\documentclass[%
 aip, cha, numerical,
 amsmath,amssymb,
reprint,%
]{revtex4-1}

\usepackage[utf8]{inputenc}
\usepackage[T1]{fontenc}
\usepackage{mathptmx}
\usepackage{color}
\usepackage{amssymb}
\usepackage{amsmath}
\usepackage{tabularx}
\usepackage{hyperref}
\usepackage{ltxtable}
\usepackage{natbib} 
\usepackage{longtable}
\usepackage{float}
\usepackage{natbib}
\usepackage{amsmath}
\usepackage{dcolumn}
\usepackage{bm}
\usepackage{graphicx}

\newcommand{\be}{\begin{equation}}
\newcommand{\ee}{\end{equation}}
\newcommand{\Hbse}{\begin{subequations}}
\newcommand{\ese}{\end{subequations}}
\newcommand{\bea}{\begin{eqnarray}}
\newcommand{\eea}{\end{eqnarray}}
\newcommand{\ba}{\begin{array}}
\newcommand{\ea}{\end{array}}
\newcommand{\bi}{\begin{itemize}}
\newcommand{\ei}{\end{itemize}}
\newcommand{\ben}{\begin{enumerate}}
\newcommand{\een}{\end{enumerate}}

\usepackage[usenames,dvipsnames]{xcolor}
\hypersetup{
    colorlinks,
    citecolor=blue,
    filecolor=blue,
    linkcolor=blue,
    urlcolor=blue}

\begin{document}

\preprint{AIP/123-QED}

\title{Kolmogorov flow: Linear Stability and Energy Transfers in a minimal low-dimensional model}

\author{Soumyadeep Chatterjee}
\email{soumyade@iitk.ac.in}
\affiliation{ 
Department of Physics, Indian Institute Of Technology Kanpur, Kanpur 208016, India
}%

\author{Mahendra K. Verma}%
\email{mkv@iitk.ac.in}
\affiliation{ 
Department of Physics, Indian Institute Of Technology Kanpur, Kanpur 208016, India
}%

\date{\today}


\begin{abstract}
    
In this paper, we derive a four-mode model for the  Kolmogorov flow by employing Galerkin truncation and  Craya-Herring basis for the decomposition of velocity field.  After this, we perform a bifurcation analysis of the model.   Though our low-dimensional model has fewer modes than the past models,  it captures the essential features of the primary bifurcation of  the Kolmogorov flow.  For example, it reproduces the critical Reynolds number for the supercritical pitchfork bifurcation and the flow structures of the past works.  We also demonstrate energy transfers from intermediate scales to large scales. We perform direct numerical simulations of the Kolmogorov flow and show that our model predictions match with the numerical simulations very well.

\end{abstract}

\keywords{Kolmogorov flow,  low-dimensional model, linear stability, bifurcation analysis{\color{blue}.} }

\maketitle


\begin{quotation}
    
In the late 1950s, Kolmogorov urged the fluid community to explore the stability criteria of a shear flow with spatially-periodic forcing; a system referred to as the Kolmogorov flow. Since then, many researchers have attempted to address the above problem using analytical, numerical, and experimental tools.  The leading analytical results involve infinite or a large number of interacting Fourier modes.   The numerical calculations and low-dimensional models too involve many Fourier modes.  In this paper, we construct a four-mode low-dimensional model using Galerkin truncation and Craya-Herring basis. Our minimal model of the Kolmogorov flow captures the essential features of its primary bifurcation and the critical Reynolds number very  well.  The model predictions are borne out in numerical simulations.

\end{quotation}


\section{INTRODUCTION}
\label{sec:level1}
Flow instability and transition to turbulence are important problems of fluid dynamics.  Kolmogorov abstracted a simple shear flow with spatially-periodic forcing~\cite{Arnold:RMS1960, Obukhov:RMS1983, Arnold:PRSL1991}, whose instability and bifurcation has been studied intensely over the years.      Also, the Kolmogorov flow has been experimentally realized in several setups, including a soap film\cite{Burgess:PRE1999} and an electrolytic fluid\cite{Bondarenko:ANS1979, Suri:POF2014}.    In this paper, we analyze the stability of the Kolmogorov flow using a low-dimensional model consisting of four Fourier modes. We validate the model using numerical solutions.

 \citet{Meshalkin:JAMM1961} provided the first solution to the stability of the Kolmogorov flow. They considered an external force per unit mass, $\gamma \sin k_f y \ \hat{x}$ where $\gamma$ is the force amplitude, and $k_f$  is the force wavenumber.   \citet{Meshalkin:JAMM1961} considered $k_f = 1$, and analyzed the stability of two-dimensional flow with $L_x/L_y=1/\alpha$.    They considered small perturbation on the fundamental stream function (corresponding to the external force) and focused on its time variations. They incorporated the effects of all Fourier modes and studied the stability problem with continued fractions. An outcome of their analysis is that for $\alpha<1$, the laminar solution becomes unstable  at critical Reynolds number $R_c$; and $R_c \rightarrow \sqrt{2}$ (for normalization of \citet{Iudovich:JAMM1965}) as $\alpha\rightarrow0$.  They observed that the laminar solution is stable for $\alpha>1$. Using asymptotic instability analysis, \citet{Sivashinsky:PD1985}  showed that a periodically-forced two-dimensional plane-parallel flow becomes unstable beyond a critical Reynolds number.  He showed the secondary flow to be chaotically self-fluctuating. 
 
\citet{Iudovich:JAMM1965} and \citet{Marchioro:CMPP1987}  extended the calculation of   \citet{Meshalkin:JAMM1961}   and  concluded that the laminar flow is globally stable for $\alpha\ge1$.     For $\alpha<1$, \citet{Iudovich:JAMM1965} proved that $R_c \rightarrow \sqrt{2}$ for $\alpha \rightarrow 0$, and $R_c \rightarrow \infty$ when $\alpha\rightarrow 1$.  They showed that $R_c$ increases monotonically with $\alpha$ between $R_c = \sqrt{2}$ for $\alpha \rightarrow 0$ and $R_c = \infty$ for $\alpha \rightarrow 1$.  The $R_c$ curve represents neutral stability.

\citet{Okamoto:JJIAM1993} performed a bifurcation analysis of the Kolmogorov flow with a finite set of Fourier modes and showed supercritical pitchfork to be the primary bifurcation.  They considered $544$ modes for  $\alpha > 0.3$,  even more modes for $\alpha < 0.3$, and observed that $R_c =3.011193$ for $\alpha =0.7$.  Using  more sophisticated calculation, \citet{Nagatou:JCAM2004} reported  $R_c$ to be bracketed  between $3.011528364444$ and $3.011528364446$.  Later, \citet{Okamoto:JDDE1996, Okamoto:DM1998} extended the bifurcation diagram to larger $R$  using path-continuation method. \citet{Matsuda:TMJ2002} studied the bifurcation diagram further and derived an exact formula for the second derivatives of their components   at the bifurcation points. 

The Kolmogorov flow has been simulated in experiments by inducing vortices in magnetofluids using periodically placed electrodes.  \citet{Tabeling:EPL1987} observed supercritical pitchfork bifurcation at the instability of the vortices.   \citet{Bondarenko:ANS1979} performed a similar experiment.  In another experiment, \citet{Sommeria:JFM1986} reported the existence of an inverse cascade due to the nonlinear interactions. \citet{Herault:EPL2015} observed $ 1/f  $ noise in the nonlinear regime of the Kolmogorov flow. 	 \citet{Tabeling:PR2002} reviewed the experiments related to the Kolmogorov flow.

\citet{Gotoh:FDR1987} performed instability analysis of the rhombic cells with the stream function as $\cos kx+\cos y$, where $k$ is the aspect ratio of the cell.  \citet{Kim:IMAJAM2003} performed bifurcation and inviscid limit analysis for  the aforementioned rhombic cells. \citet{Thess:POFA1992}  studied the effects of viscosity, linear friction and confinement on the flow.  \citet{Platt:PFA1991} analyzed the Kolmogorov flow for $k_f = 4$ and observed a sequence of bifurcations leading to chaos. For the same $k_f$, \citet{Chen:Chaos2004} studied the chaotic behavior using a truncation model with nine modes.   In addition, researchers have studied variations of the Kolmogorov flow to three-dimensional flows \cite{Sarris:POF2007}.        

 
The Kolmogorov flow is useful not only for analyzing transition to turbulence but also for studying the inverse cascade in two-dimensional turbulence\cite{Green:JFM1974, Gupta:PRE2019, Zhang:POF2019}. 
\citet{Green:JFM1974} reported that for $k>k_f$, kinetic energy spectrum, $E(k)\sim k^{-5}$ whereas for $k<k_f$, $E(k)\sim k$. \citet{Sommeria:JFM1986} studied the inverse cascade experimentally and reported that the exponent to be in the range of $-4.5$ to $-4.9$ for $k > k_f$. For $k  < k_f$,  direct numerical simulations (DNS) reveal that $E(k)\sim k^{-5/3}$.  For random forcing in a wavenumber band near $k=k_f$, \citet{Gupta:PRE2019}  showed that for $k>k_f$, the energy spectrum is of the form $k^{-3} \exp(-k^2)$; the exponential part gives an appearance of steeper spectrum compared to $k^{-3}$.   \citet{Zhang:POF2019} performed a molecular  simulation using Fokker-Planck method and reported that $E(k)\sim k^{-4}$ for $k < k_f$ due to condensation in the large scale structures.   For $ k> k_f$,   \citet{Zhang:POF2019} reported that $ E(k) \sim \exp(-0.2 k) $. Energy condensate is observed at the large-scale due to inverse cascade. \citet{Gallet:JFM2013} derived a mathematical model of energy condensation in the absence of large-scale dissipation.  \citet{Mishra:PRE2015}  studied the condensate regime using Ekman friction.  There are more works on the Kolmogorov flow, including those by \citet{Chandler:JFM2013}, \citet{Lucas:JFM2014} \& \citet{Fylladitakis:JAMP2018}, and  references therein.

In this paper, we consider incompressible Kolmogorov flow in a two-dimensional periodic box with aspect ratio $\alpha$, and construct a low-dimensional model with four modes.  We perform bifurcation analysis of the system and derive the critical Reynolds number for the instability.  Our results are consistent with previous ones. Besides, we also carry out direct numerical simulations of the Kolmogorov flow  for the parameter used for our model.  The results from these simulations are in good agreement with those from the low-dimensional model.  These results provide us confidence that the chosen modes are a good choice for the Kolmogorov flow.

The outline of this paper is as follows. In Sec.~\ref{sec:level2}, we present the governing equations and low-dimensional model.   In Sec.~\ref{sec:level3}, we perform linear stability and bifurcation analysis of the low-dimensional model.  We describe the energy transfers among the participating modes in Sec.~\ref{sec:level4}. In Sec.~\ref{sec:level5} we present numerical validation using direct numerical simulation.We conclude in Sec.~\ref{sec:level6}. 


\section{Basic formulation - }\label{sec:level2}
For an incompressible flow, the Navier-Stokes equation and incompressibility condition \citep{Verma:book:BDF, Verma:book:ET} are
\bea
\frac{\partial \mathbf{u}}{\partial t}+\mathbf{u} \cdot \mathbf{\nabla}\mathbf{u} & = &  -\mathbf{\nabla}p + \mathbf{F}_{u}+ \nu \nabla^{2} \mathbf{u}, \label{eq:eom} \\  \mathbf{\nabla}\cdot \mathbf{u} & = &  0, \label{eq:incompres} 
\eea
where $\mathbf{u}$ and $p$ are the velocity and pressure fields respectively, $\nu$ is the kinematic viscosity, and $\mathbf{F}_{u} $ is the acceleration due to the external force. We consider two-dimensional Kolmogorov flow in a doubly-periodic box of size $L_x \times L_y$. The ratio $\alpha = L_y/L_x$ is called aspect ratio. We assume density $\rho$ to be unity. We take
\be
\mathbf{F}_{u} = \gamma \sin \bigg(\frac{2\pi y k_{f}}{L_{y}}\bigg) \hat{x}, \label{eq:forcing}   
\ee
where $\gamma$ is the amplitude of the acceleration, and $k_f$ is the forcing wavenumber which we consider to be $1$.    

We nondimensionalize Eqs.~(\ref{eq:eom}, \ref{eq:incompres}) using ${L_{y}}/{2\pi}$ as the length scale and  ${2\pi\nu}/{\gamma L_{y}}$ as the time scale, and obtain the  following equations:
\bea
\frac{\partial \mathbf{u}}{\partial t}+\mathbf{u} \cdot \mathbf{\nabla}\mathbf{u} & = & -\mathbf{\nabla}p+\frac{1}{R}\sin(y)\hat{x}+ \frac{1}{R} \nabla^{2} \mathbf{u}, \label{eq:eom1} \\  \mathbf{\nabla}\cdot \mathbf{u} & = & 0,\label{eq:incompres1} 
\eea 
where
\be
R=\frac{\gamma}{\nu^{2}}\bigg(\frac{L_{y}}{2\pi}\bigg
)^3,
\ee
is the Reynolds number. A trivial stationary solution of the Eqs.~(\ref{eq:eom1}, \ref{eq:incompres1}) is  
\be
 \mathbf{u} \ = \ (\sin(y),0), \quad p=\mathrm{const.}\label{meanflow}   
\ee
This solution represents a laminar flow.

For stability analysis, it is customary to work in Fourier space. In this space, the equations get transformed to
\bea
\frac{d}{dt} \mathbf{u}(\mathbf{k})+\mathbf{N_{u}}(\mathbf{k}) & = & -i\mathbf{k}p(\mathbf{k})+\mathbf{F}_{u}(\mathbf{k})-\frac{1}{R} k^{2}\mathbf{u}(\mathbf{k}),\label{eq:eom2}\\
\mathbf{k}.\mathbf{u}(\mathbf{k}) & = & 0, \label{eq:incompress2}   
\eea   
where
\bea
\mathbf{N_{u}}(\mathbf{k}) & = &  i\sum\limits_{\mathbf{p}}\{\mathbf{k}\cdot\mathbf{u}(-\mathbf{q})\}\mathbf{u}(-\mathbf{p}),\label{eq:nonlin}  \\
p(\mathbf{k}) & = & \frac{i}{k^2} {\bf k} \cdot [\mathbf{N_{u}}(\mathbf{k})  
- \mathbf{F_{u}}(\mathbf{k}) ],
\eea
are the Fourier transforms of the nonlinear and pressure terms respectively. Here ${\bf k = -p -q}$. Note that calculation of $\mathbf{N_{u}}(\mathbf{k})$ requires all possible wavenumber triads.  We denote the  Fourier mode as ${\bf k} = (\alpha l, m)$, where $l,m$ are integers. 

\begin{figure}[htbp]
    \centering
    \includegraphics[width=1.0\linewidth]{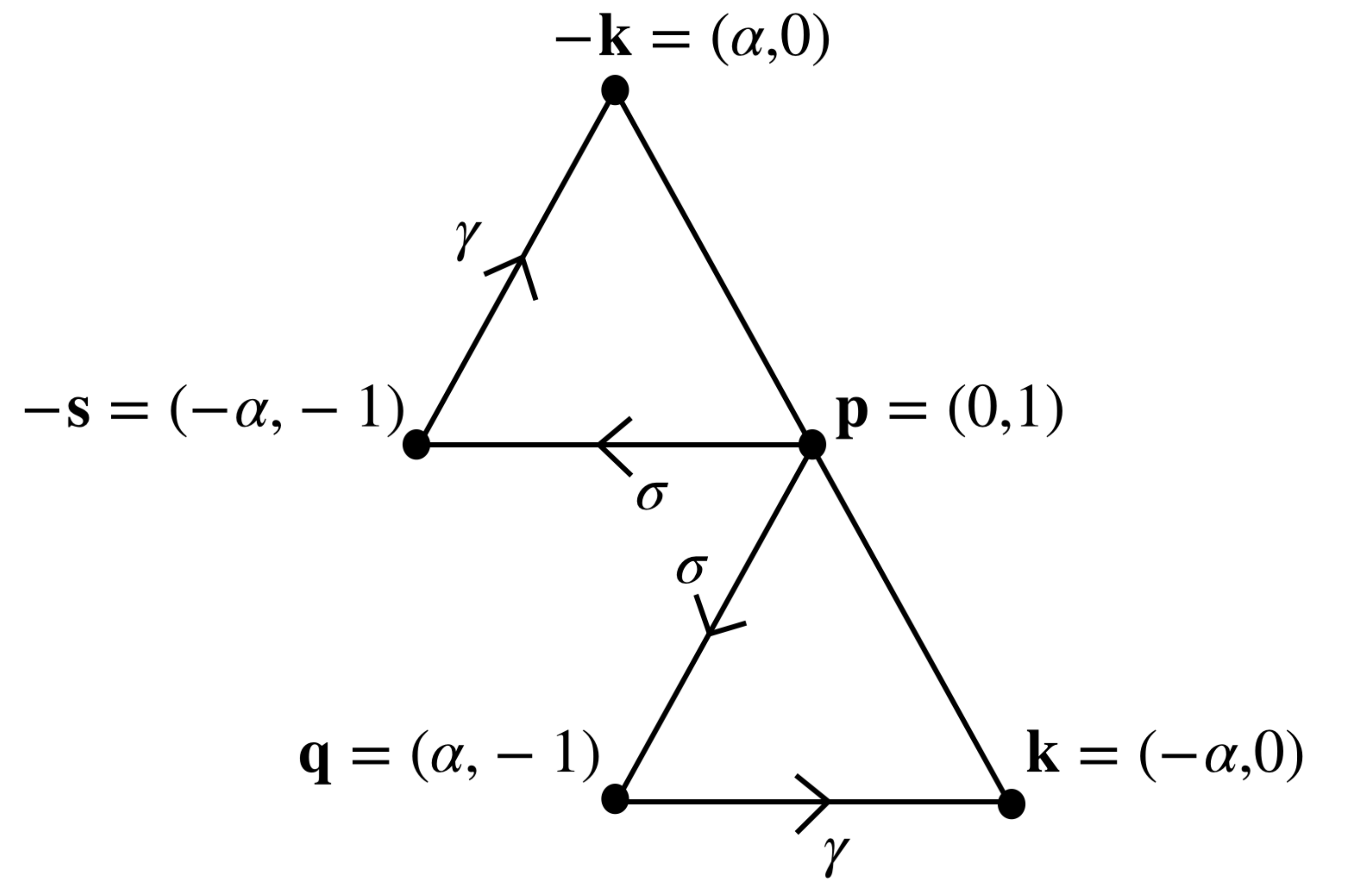}
    \caption{Schematic representation of the triad interactions in Eqs. ~(\ref{triad_1},\ref{triad_2}). Energy transfers $\sigma$ and $\gamma$ are $(r-1)/(4\sqrt{2}Rr^{2})$, $(\alpha^{2}(r-1))/(4\sqrt{2}Rr^{2})$, where $r=R/R_c$.}
    \label{fig:triad_diagram}
\end{figure}

The Fourier transform of the external force $\frac{1}{R}\sin(y)\hat{x}$ is
\be
\mathbf{F}_{u}(\mathbf{k})  =  \frac{1}{2iR}[\delta_{k_{y},1}-\delta_{k_{y}, -1}]\hat{x}.\label{eq:force}
\ee
That is, the forcing wavenumbers are $(0,1)$ and $(0,-1)$. Near the onset of instability, the nonlinear term ${\bf u \cdot \nabla u}$ generates other Fourier modes. In this paper we show that a low-dimensional model having nonzero Fourier modes at wavenumbers $\{ \mathbf{k} \ = \ (-\alpha,0)$, $\mathbf{p} \ = \ (0,1)$, $\mathbf{q} =  (\alpha,-1)$, and $\mathbf{s} \ = \ (\alpha,1) \}$ reproduces earlier results on Kolmogorov flow quite well (e.g. \citet{Iudovich:JAMM1965}).  In Sec.~\ref{sec:level5} we perform direct numerical numerical simulations and show that our low-dimensional model reproduces  the simulation results to a significant degree. Due to these reasons, we work with this set of Fourier modes. We consider the following interacting triads:
\bea
{\bf k} \bigoplus {\bf p} \bigoplus  {\bf q} = (-\alpha,0)\bigoplus(0,1)\bigoplus(\alpha,-1) & = & 0, \label{triad_1} \nonumber  \\ \\
({\bf -k}) \bigoplus {\bf p} \bigoplus ( {\bf -s} ) = (\alpha,0)\bigoplus(0,1)\bigoplus(-\alpha,-1) & = & 0, \label{triad_2} \nonumber  \\
\eea
where $\bigoplus$ represents nonlinear interaction (see Fig.~\ref{fig:triad_diagram}).
Thus possible nonlinear interactions for the modes with wavenumbers $\bf{k}$, $\bf{p}$, $\bf{q}$ and $\bf{s}$ are
\bea
    \mathbf{k} & = & [(-\alpha,-1)\bigoplus(0,1)]+[(-\alpha,1)\bigoplus(0,-1)], \label{eq:com1}\\
    \mathbf{p} & = & [(-\alpha,0)\bigoplus(\alpha,1)]+[(\alpha,0)\bigoplus(-\alpha,1)], \label{eq:com2}\\
    \mathbf{q} & = & [(\alpha,0)\bigoplus(0,-1)], \label{eq:com3}\\
    \mathbf{s} & = & [(\alpha,0)\bigoplus(0,1)].  \label{eq:com4}
\eea

In the next section we derive the governing equations for the above Fourier modes.


\section{Linear stability and bifurcation analysis}\label{sec:level3}

The derivation of the evolution equations for the Fourier modes get simplified in Craya-Herring basis~\cite{Craya:thesis, Herring:PF1974, Lesieur:book:Turbulence, Sagaut:book, Verma:book:ET}.  For wavenumber ${\bf k}$, the unit vectors in this basis are~\cite{Craya:thesis, Herring:PF1974, Lesieur:book:Turbulence, Sagaut:book, Verma:book:ET}
\bea
\hat{e}_{3}(\mathbf{k}) & = & \hat{k}, \label{eq:e3}\\
 \hat{e}_{1}(\mathbf{k}) & = & \frac{\hat{k}\times\hat{n}}{|\hat{k}\times\hat{n}|}, \label{eq:e1}\\
\hat{e}_{2}(\mathbf{k}) & = & \hat{e}_{3}(\mathbf{k})\times \hat{e}_{1}(\mathbf{k}), \label{eq:e2}
\eea
where $\hat{k}$ is the unit vector along $\mathbf{k}$,  and  $\hat{n} = \hat{z}$. For all the wavenumbers under consideration, $\hat{e}_{1}$ lie on the plane, while $\hat{e}_{2} $ are perpendicular to the plane. Hence, for the present 2D flow, $u_2 = 0$ for all the modes. In addition, $u_3 = 0$ due to incompressibility condition (Eq.~(\ref{eq:incompress2})). Therefore, 
\be
\mathbf{u}({\mathbf{k}}) = u_{1}(\mathbf{k})\hat{e}_{1}(\mathbf{k})\label{CHvelocity}.
\ee
Explicitly, the unit vectors $\hat{e}_{1}$'s for the four wavenumbers ($\mathbf{k}$,$\mathbf{p}$, $\mathbf{q}$, $\mathbf{s}$) are
\bea
    \hat{e}_{1}(\mathbf{k}) & = & \hat{y},\label{e1k} \\
    \hat{e}_{1}(\mathbf{p}) & = & \hat{x},\label{e1p} \\
    \hat{e}_{1}(\mathbf{q}) & = & -\frac{1}{\sqrt{\alpha^{2}+1}}\hat{x}-\frac{\alpha}{\sqrt{\alpha^{2}+1}}\hat{y},\label{e1q}\\
    \hat{e}_{1}(\mathbf{s}) & = & \frac{1}{\sqrt{\alpha^{2}+1}}\hat{x}-\frac{\alpha}{\sqrt{\alpha^{2}+1}}\hat{y}.\label{e1s}
\eea

Using Eqs.~(\ref{e1k}-\ref{e1s}), we derive the  evolution equations for   $u_1$'s as
\bea
\frac{d}{dt}u_{1}(\mathbf{k})  & = &  -\bigg( \frac{\alpha^{2}}{\sqrt{1+\alpha^{2}}}i\bigg) (u^{*}_{1}(\mathbf{p})u^{*}_{1}(\mathbf{q})+u^{*}_{1}(\mathbf{s})u_{1}(\mathbf{p}))\nonumber\\&& -  \bigg(\frac{\alpha^{2}}{R}\bigg) u_{1}(\mathbf{k}), \label{eq:eomk}\\
\frac{d}{dt}u_{1}(\mathbf{p})  & = &  \bigg( \frac{1}{\sqrt{1+\alpha^{2}}}i\bigg) (u^{*}_{1}(\mathbf{k})u^{*}_{1}(\mathbf{q})-u_{1}(\mathbf{k})u_{1}(\mathbf{s}))\nonumber\\ &&+ \frac{1}{2iR}-\bigg(\frac{1}{R}\bigg) u_{1}(\mathbf{p}),\label{eq:eomp}\\
\frac{d}{dt}u_{1}(\mathbf{q})  & = & -\bigg( \frac{1-\alpha^{2}}{\sqrt{1+\alpha^{2}}}i\bigg) u^{*}_{1}(\mathbf{k})u^{*}_{1}(\mathbf{p})
-\bigg(\frac{1+\alpha^{2}}{R}\bigg)u_{1}(\mathbf{q}),\label{eq:eomq} \nonumber \\ \\
\frac{d}{dt}u_{1}(\mathbf{s})  & = &  -\bigg( \frac{1-\alpha^{2}}{\sqrt{1+\alpha^{2}}}i\bigg) u^{*}_{1}(\mathbf{k})u_{1}(\mathbf{p})
-\bigg(\frac{1+\alpha^{2}}{R}\bigg)u_{1}(\mathbf{s}).\label{eq:eoms} \nonumber \\ 
\eea
The steady-state solutions of the above equations are
\bea
S_0: 
\begin{cases}
     u_{1}(\mathbf{k})= 0,\\
    u_{1}(\mathbf{p}) = \frac{1}{2i},\\
    u_{1}(\mathbf{q}) =0,\\
    u_{1}(\mathbf{s}) =0 \label{eq:baseflow},
\end{cases}
\eea
\bea
S_1:
\begin{cases}
     u_1(\mathbf{k})  =  -\frac{1}{\sqrt{2}r}\sqrt{r-1},\\
    u_1(\mathbf{p})  =  -\frac{1}{2r}i,\\ 
    u_1(\mathbf{q})  =  -\frac{\sqrt{1+\alpha^{2}}}{2R}\sqrt{r-1},\\ 
    u_1(\mathbf{s})   =  \frac{\sqrt{1+\alpha^{2}}}{2R}\sqrt{r-1}\label{eq:possible1},  
\end{cases} \eea
and 
\bea
S_2: 
\begin{cases}
    u_1(\mathbf{k})  =  \frac{1}{\sqrt{2}r}\sqrt{r-1},\\
    u_1(\mathbf{p})  =  -\frac{1}{2r}i,\\ 
    u_1(\mathbf{q})  =  \frac{\sqrt{1+\alpha^{2}}}{2R}\sqrt{r-1},\\ 
    u_1(\mathbf{s})   =  -\frac{\sqrt{1+\alpha^{2}}}{2R}\sqrt{r-1},\label{eq:possible2}
\end{cases}\eea 
where $r=R/R_c$ with
\be 
R_{c} \ = \ \frac{\sqrt{2}(1+\alpha^{2})}{\sqrt{1-\alpha^{2}}}.  \label{Critical_Reynolds_number}
\ee

\begin{figure}[htbp]
    \centering
    \includegraphics[width=0.97\linewidth]{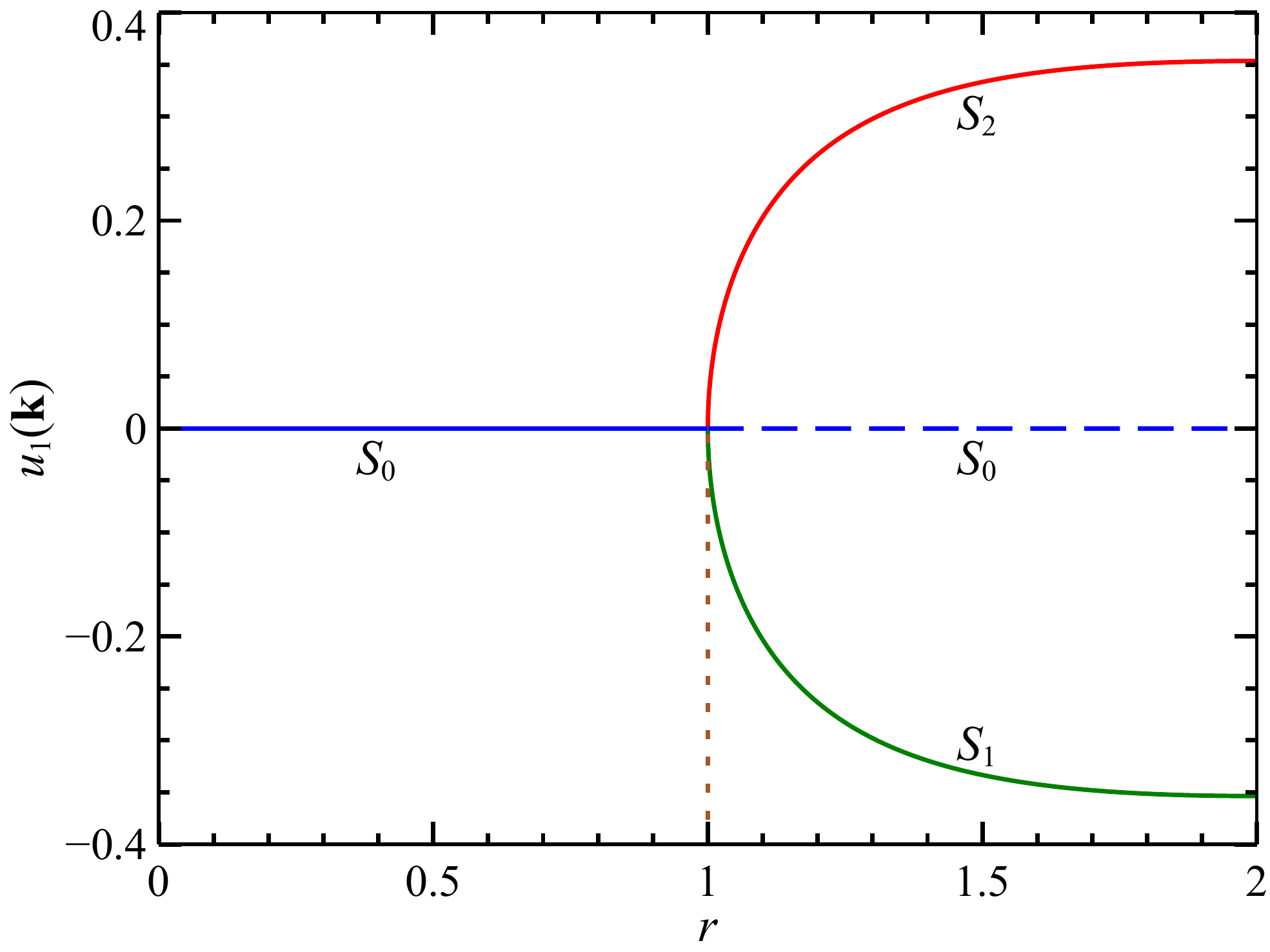}
    \caption{(color online) For $\alpha \ = \ 0.7$ ,   $u_{1}(\mathbf{k})$ as a function of $r$.}
    \label{fig:bifuraction} 
\end{figure}

The solution $S_0$ is valid for all $r$, while $S_1$ and $S_2$ are defined only for $r > 1$.  Also, $u_1(\mathbf{k}), u_1(\mathbf{q}), u_1(\mathbf{s})$ modes of $S_1$ have opposite signs compared to $S_2$.  See Fig.~\ref{fig:bifuraction}  for an  illustration of steady  $u_1(\mathbf{k})$ for $\alpha=0.7$; the figure exhibits a transition from $S_0$ to $S_1$ or $S_2$  at $r =1$.  For $r>1$, the system  follows either $S_1$ branch or $S_2$ branch  depending on the  initial condition.  In subsequent subsections, we show  that $S_0$ is the only stable solution for $r<1$, while $S_1$ and $S_2$ are the stable solutions for $r>1$.   For $r>1$, the solution $S_0$ is unstable. Another important point to note that $R_c, S_1, S_2$ are not defined for $\alpha >1$, hence $S_0$ is the only solution for $\alpha > 1$. 

In the following subsections, we analyze the stability of $S_0$, $S_1$, and $S_2$.


\subsection{Stability of laminar solution $S_0$}\label{subsec:level3a}

First, we analyse the  stability of the laminar solution, $S_0$.   For the same, we linearize Eqs.~(\ref{eq:eomk}-\ref{eq:eoms}) around $S_0$ and obtain the following equations: 
 \bea
\frac{d}{dt}\tilde{u}_{1}(\mathbf{k})  &= &  -\bigg(\frac{\alpha^{2}}{2\sqrt{1+\alpha^{2}}}\bigg) (\tilde{u}^{*}_{1}(\mathbf{s})-\tilde{u}^{*}_{1}(\mathbf{q})) -\bigg(\frac{\alpha^{2}}{R}\bigg) \tilde{u}_{1}(\mathbf{k}), \label{eq:eomk1} \nonumber \\ \\
\frac{d}{dt}\tilde{u}_{1}(\mathbf{p}) & = & - \frac{1}{R} \tilde{u}_{1}(\mathbf{p}), \label{eq:eomp1}\\
\frac{d}{dt}\tilde{u}_{1}(\mathbf{q})  &=  &\bigg( \frac{(1-\alpha^{2})}{2\sqrt{1+\alpha^{2}}}\bigg) \tilde{u}^{*}_{1}(\mathbf{k})-\bigg(\frac{1+\alpha^{2}}{R}\bigg)\tilde{u}_{1}(\mathbf{q}),\label{eq:eomq1}\\
\frac{d}{dt}\tilde{u}_{1}(\mathbf{s})  &= & -\bigg( \frac{(1-\alpha^{2})}{2\sqrt{1+\alpha^{2}}}\bigg) \tilde{u}^{*}_{1}(\mathbf{k})-\bigg(\frac{1+\alpha^{2}}{R}\bigg)\tilde{u}_{1}(\mathbf{s}),
\label{eq:eoms1}
\eea  
where $\tilde{u}_{1}(\mathbf{k})$, $\tilde{u}_{1}(\mathbf{p})$, $\tilde{u}_{1}(\mathbf{q})$ and $\tilde{u}_{1}(\mathbf{s})$ represent fluctuations in $S_0$, and they are  complex quantities. Hence we split them into real and imaginary parts:
\bea
     \tilde{u}_{1}(\mathbf{k}) & = & \Re[\tilde{u}_{1}(\mathbf{k})]+ i \Im[\tilde{u}_{1}(\mathbf{k})],\\
     \tilde{u}_{1}(\mathbf{p}) & = & \Re[\tilde{u}_{1}(\mathbf{p})]+ i \Im[\tilde{u}_{1}(\mathbf{p})],\\
     \tilde{u}_{1}(\mathbf{q}) & = & \Re[\tilde{u}_{1}(\mathbf{q})]+ i \Im[\tilde{u}_{1}(\mathbf{q})],\\
     \tilde{u}_{1}(\mathbf{s}) & = & \Re[\tilde{u}_{1}(\mathbf{s})]+ i \Im[\tilde{u}_{1}(\mathbf{s})].
\eea
Using Eqs.~(\ref{eq:eomk1}-\ref{eq:eoms1}) we derive the following matrix equation:
\be
    \frac{d}{dt}\mathbb{U}=\mathbb{A}\mathbb{U},\label{eq:matrixequation}
\ee
where 
\bea
\mathbb{A}&=&\begin{pmatrix}
    -\frac{A}{R} & 0 & 0 & 0 & B & 0 & -B & 0 \\
    0 & -\frac{A}{R} & 0 & 0 & 0 & -B & 0 & B \\
    0 & 0 & -\frac{1}{R} & 0 & 0 & 0 & 0 & 0 \\
    0 & 0 & 0 & -\frac{1}{R} & 0 & 0 & 0 & 0 \\
    D & 0 & 0 & 0 & -\frac{C}{\mathrm{R}} & 0 & 0 & 0\\
    0 & -D & 0 & 0 & 0 & -\frac{C}{\mathrm{R}} & 0 & 0 \\
    -D & 0 & 0 & 0 & 0 & 0 & -\frac{C}{\mathrm{R}} & 0 \\
    0 & D & 0 & 0 & 0 & 0 & 0 & -\frac{C}{\mathrm{R}} 
\end{pmatrix},\nonumber\\
\mathbb{U}&=&\begin{pmatrix}
    \Re[\tilde{u}_{1}(\mathbf{k})] \\
    \Im[\tilde{u}_{1}(\mathbf{k})]\\
    \Re[\tilde{u}_{1}(\mathbf{p})] \\
    \Im[\tilde{u}_{1}(\mathbf{p})]\\
    \Re[\tilde{u}_{1}(\mathbf{q})]\\
    \Im[\tilde{u}_{1}(\mathbf{q})]\\
    \Re[\tilde{u}_{1}(\mathbf{s})]\\
    \Im[\tilde{u}_{1}(\mathbf{s})]
\end{pmatrix}.
\eea
Here, $\mathbb{A}$ is a $8\times8$  matrix with $A = \alpha^{2}$, $B = \alpha^{2}/(2\sqrt{1+\alpha^{2}})$, $C = 1+\alpha^{2}$ and $D = (1-\alpha^{2})/(2\sqrt{1+\alpha^{2}})$.  The solution $\mathbb{U}(t)$ is a linear combination of $e^{\lambda t}$, where $\lambda$'s are the eigenvalues of  $\mathbb{A}$:
\bea
\lambda_{1} & = & -\frac{1}{R},\\
\lambda_{2} & = & -\frac{C}{R},\\
\lambda_{3} & = & -\frac{A+C+\sqrt{(A-C)^{2}+8BDR^{2}}}{2R}, \\ 
\lambda_{4} & = & -\frac{A+C-\sqrt{(A-C)^{2}+8BDR^{2}}}{2R}.
\eea 
It is easy to show that $\lambda_{1}$, $\lambda_{2}$, and $\lambda_{3}$ are negative for all $\alpha$ and ${R}$. However, $\lambda_{4}$ changes sign from negative to positive at the following condition, called {\em neutral stability condition}~\cite{Chandrasekhar:book:Instability}:
\be
 \sqrt{(A-C)^{2}+8BD\mathrm{R}^{2}} = A+C, ~~~\mathrm{or}~~R = R_c, \label{eq:insta}  
\ee
where $R_c$ is given by Eq.~(\ref{Critical_Reynolds_number}).

\begin{figure}[htbp]
    \centering
    \includegraphics[width=0.97 \linewidth]{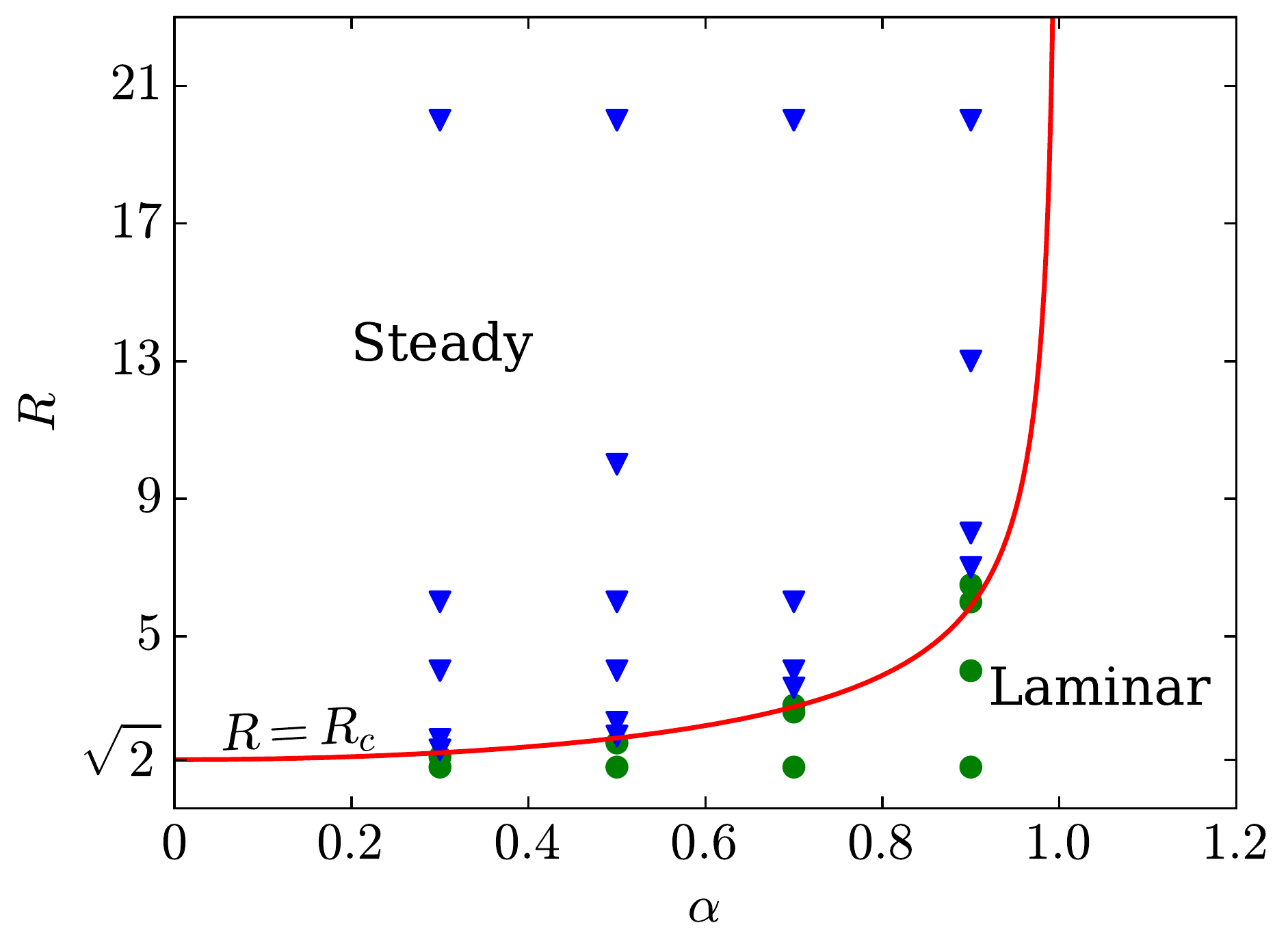}
    \caption{(color online) Phase diagram $(R, \alpha)$ constructed using $R= R_c$ curve, also called neutral stability curve (the red curve).  The figure also includes the steady-state results of the DNS.  The green circles represent laminar solution ($ S_0 $), but the blue triangles represent the steady vortex flow ($ S_1$ or $S_2 $).}  
    \label{fig:neutral_curve} 
\end{figure}

In Fig.~\ref{fig:neutral_curve} we exhibit the $(R, \alpha)$ phase diagram, the $R_c$ curve, as well as  the regions of stability and instability. As shown in the figure, $R_c$ increases monotonically with $\alpha$, and $R_c \rightarrow \infty$ as $\alpha \rightarrow 1$. Also $R_c \rightarrow \sqrt{2}$ as $\alpha \rightarrow 0$.   The figure shows that the system is stable below the $R=R_c$ curve and yields the laminar solution ($ S_0 $), and unstable otherwise.   Note that for $\alpha > 1$, the laminar solution is stable for all $R$.  
 
In the next section we show that for $\alpha < 1$ and $R> R_c$, the  solutions $S_1$ and $S_2$ are the stable solutions.  


\subsection{Stability analysis of $S_1$ and $S_2$}\label{subsec:level3b}

As described in Sec.~\ref{sec:level3}, the solutions  $S_1, S_2$ exist only for $r > 1$. In this subsection we show that these are stable solutions for $r>1$.   For the stability analysis, we  generate the stability matrix for $S_1$ and $S_2$ by linearizing  Eqs.~(\ref{eq:eomk}-\ref{eq:eoms}) around $S_1, S_2$.  This exercise yields a set of equations for $\tilde{u}_{1}(\bf{k})$, $\tilde{u}_{1}(\bf{p})$, $\tilde{u}_{1}(\bf{q})$, and $\tilde{u}_{1}(\bf{s})$ similar to Eqs.~(\ref{eq:eomk1}-\ref{eq:eoms1}). Note that $\tilde{u}_{1}(\bf{k})$, $\tilde{u}_{1}(\bf{p})$, $\tilde{u}_{1}(\bf{q})$, and $\tilde{u}_{1}(\bf{s})$ are fluctuations around $S_1,S_2$. The resulting matrix equation is
\be
\frac{d}{dt}\mathbb{U}=\mathbb{B}\mathbb{U},\label{eq:matrixequation1}
\ee
where 
\be
    \mathbb{B}=\begin{pmatrix}
    -\frac{A}{R} & 0 & 0 & \pm C' & D' & 0 & -D' & 0 \\
    0 & -\frac{A}{R} & 0 & 0 & 0 & -D' & 0 & D' \\
    0 & 0 & -\frac{1}{R} & 0 & 0 & \pm E' & 0 & \pm E' \\
    \mp H' & 0 & 0 & -\frac{1}{R} & \pm E' & 0 & \mp E' & 0 \\
    F' & 0 & 0 & \pm G' & -\frac{C}{\mathrm{R}} & 0 & 0 & 0\\
    0 & -F' & \pm G' & 0 & 0 & -\frac{C}{\mathrm{R}} & 0 & 0\\
    -F' & 0 & 0 & \mp G' & 0 & 0 & -\frac{C}{\mathrm{R}} & 0 \\
    0 & F'  & \pm G' & 0 & 0 & 0 & 0 & -\frac{C}{\mathrm{R}}
    \end{pmatrix},
\ee
and $C'$,$D'$,$E'$,$F'$,$G'$ and $H'$ are $(A A')/R$, $(B R_c)/R$, $-(B A' R_c)/(AR)$, $(D R_c)/2R$, $(D A' R_c)/2R$ and $A'/R$ respectively. Here, $A'=\sqrt{(R/R_c)-1}$, and $A$, $B$, $C$, $D$ are same as those defined in Sec.~\ref{subsec:level3a}.

Similar to the stability analysis for $S_0$, we compute the eigenvalues of the matrix $\mathbb{B}$, which  are
\bea
\lambda_{1} & = & -\frac{C}{R},\\
\lambda_{2} & = & -\frac{1+C+\sqrt{A+8 E' G' R^{2}}}{2 R},\\
\lambda_{3} & = & -\frac{1+C-\sqrt{A+8 E' G' R^{2}}}{2 R}, \\ 
\lambda_{4} & = & -\frac{A+C+\sqrt{1+8 D' F' R^{2}}}{2 R}, \\
\lambda_{5} & = & -\frac{A+C-\sqrt{1+8 D' F' R^{2}}}{2 R}, \\
\lambda_{6} & = & -\frac{2C}{3R}+\frac{2^{\frac{4}{3}}I' R}{6O}+\frac{2^{\frac{2}{3}}O}{6R^{3}},\\
\lambda_{7} & = & -\frac{2C}{3R}-\frac{2^{\frac{4}{3}}I' R(1+\sqrt{3} i)}{12O}\nonumber\\
&&-\frac{2^{\frac{2}{3}} O (1-\sqrt{3} i)}{12 R^{3}},\\
\lambda_{8} & = & -\frac{2C}{3R}-\frac{2^{\frac{4}{3}}I' R(1-\sqrt{3} i)}{12O}\nonumber\\
&&-\frac{2^{\frac{2}{3}} O (1+\sqrt{3} i)}{12 R^{3}}.
\eea 
In the above expressions, $I'$ and $O$ are complicated functions of $\alpha$ and $R$, hence they are not presented here. The eigenvalues $\lambda_1, \lambda_{2},\lambda_{3},\lambda_{4},\lambda_{5} $ are always real and negative. However, $\lambda_{6}$, $\lambda_{7}$, $\lambda_{8}$ could become complex, still their real parts are always negative. Thus, we demonstrate that the solutions $S_1, S_2$ are stable for $r>1$ for which these solutions are defined. Using these observations, we also conclude that the transition at $r =1$ (or $R= R_c$) follows a supercritical pitchfork bifurcation, as illustrated in Fig.~\ref{fig:bifuraction} for $\alpha = 0.7$.   

\begin{figure}[htbp]
    \centering
    \includegraphics[width=0.9 \linewidth]{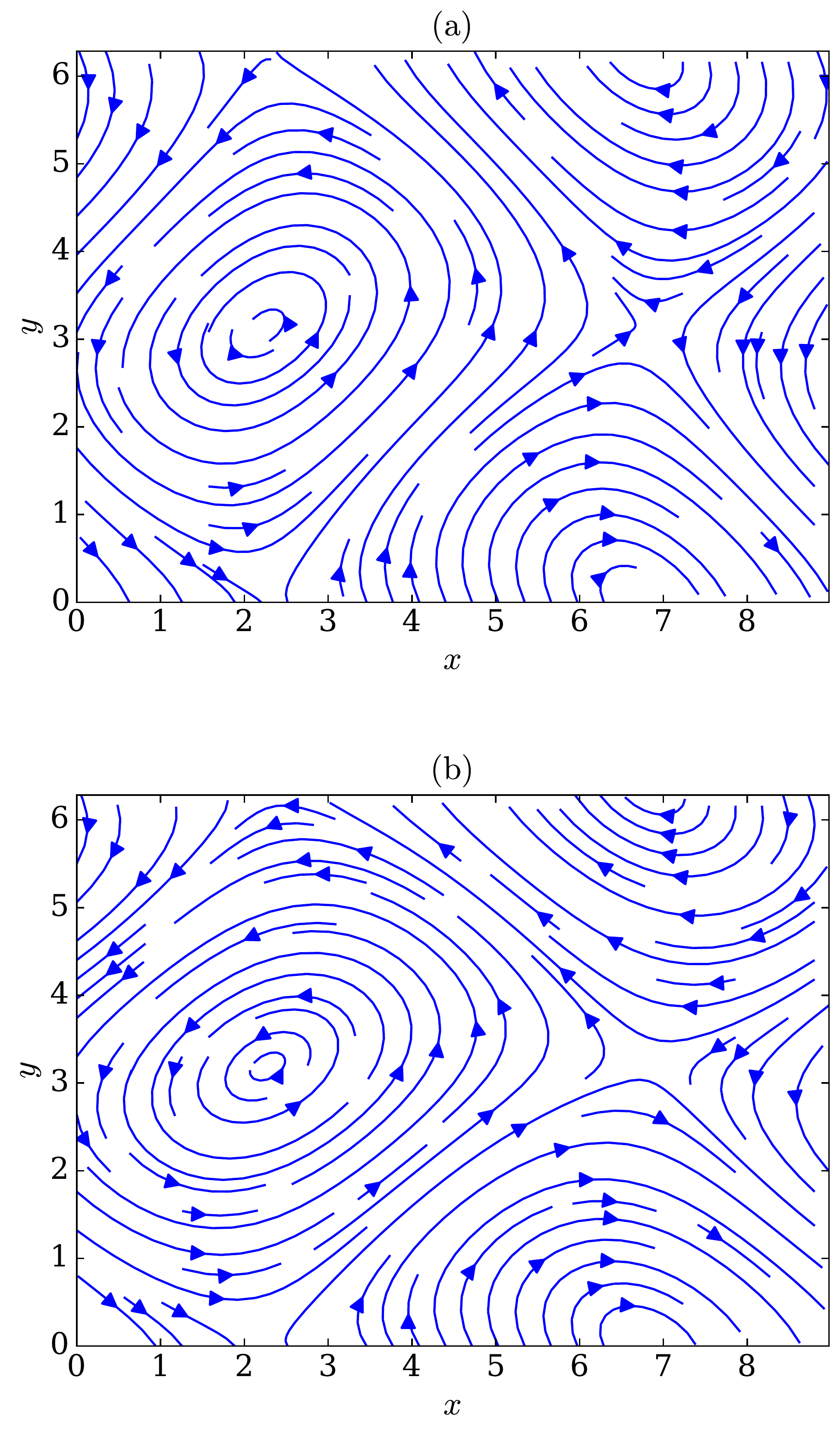}
    \caption{(color online) (a) The flow patterns  for the model solution $ S_1 $ with    $\alpha=0.7$ and $R=4.61$.  Here, $L_x = 2 \pi/\alpha$ and $L_y = 2 \pi$. (b) The same steady flow pattern, is observed in  DNS for the same parameters.}
    \label{fig:flow_profile}
\end{figure}

In Fig.~\ref{fig:flow_profile} we illustrate the flow profile of the vortex pattern generated by the modes of $S_1$ for $\alpha=0.7$ and $R=4.61$. Note that the the large scale vortical flow results from the presence of all the four modes of the model.  Interestingly, the vortical flow pattern  is very  similar to those presented in earlier works, e.g. Okamoto and Shoji\cite{Okamoto:JJIAM1993}.  Note that $R_c= 2.95$ for $\alpha=0.7$. 

The bifurcation mentioned above indicates that for $\alpha < 1$, Fourier modes with large wavelengths get excited.  In the next section, we will describe energy transfers among the interacting Fourier modes.

\section{Energy transfers in the Kolmogorov flow}\label{sec:level4}
In this section we will quantify the energy transfers between the interacting Fourier modes $\mathbf{u(k)}$, $\mathbf{u(p)}$, $\mathbf{u(q)}$ and $\mathbf{u(s)}$.  For the same we will employ the mode-to-mode energy transfer  formalism proposed by \citet{Dar:PD2001} and \citet{Verma:PR2004}. For  triad (${\bf k,p,q}$) satisfying ${\bf k+p+q}=0$,  the energy transfer from  $\mathbf{u}(\mathbf{p})$ to  $\mathbf{u}(\mathbf{k})$ with the mediation of  $\mathbf{u}(\mathbf{q})$ is  
\be
\mathit{S}^{uu}(\mathbf{k}|\mathbf{p}|\mathbf{q})= -\Im[\{\mathbf{k}\cdot\mathbf{u}(\mathbf{q})\}\{\mathbf{u}(\mathbf{p})\cdot \mathbf{u}(\mathbf{k})\}].
\ee    
For the transfers between  other Fourier modes, we employ the corresponding giver, receiver and mediator Fourier modes.

For the laminar solution $S_0$, energy transfers among the Fourier modes vanish due to a lack of nonzero interacting triad.   For $S_1, S_2$, there is no energy exchange between the Fourier modes $\mathbf{u(k)}$ and $\mathbf{u(p)}$, that is, 
\be
\mathit{S}^{uu}(\mathbf{k}|\mathbf{p}|\mathbf{q})=0.
\ee
 However, there are energy transfers among other Fourier modes.  They are,
\bea
\mathit{S}^{uu}(\mathbf{k}|\mathbf{q}|\mathbf{p})  =  \mathit{S}^{uu}(\mathbf{-k}|\mathbf{-s}|\mathbf{p}) & = & \gamma = \frac{\alpha^{2}(r-1)}{4\sqrt{2}Rr^{2}}, \\
\mathit{S}^{uu}(\mathbf{q}|\mathbf{p}|\mathbf{k})  = \mathit{S}^{uu}(\mathbf{-s}|\mathbf{p}|\mathbf{-k}) & = & \sigma = \frac{(r-1)}{4\sqrt{2}Rr^{2}}.
\eea  
 It is evident that $\gamma$ and $\sigma$ are positive because $r>1$. Also $\sigma > \gamma$ because  $\alpha < 1$.  These energy transfers are illustrated in Fig.~\ref{fig:triad_diagram}.   
 
 The energy transfer computations indicate that the velocity mode ${\bf u}(0,1)$ gives energy to ${\bf u}(\alpha,-1)$, which in turn gives energy to ${\bf u}(\alpha,0)$.  Since $\alpha < 1$, the wavenumber $(\alpha,0)$  yields the largest wavelength.  Thus,   the energy flows from intermediate scale (corresponding to wavenumber (0,1)) to large scale (coresponding to wavenumber $(\alpha,0)$).  Hence, we conclude that the Kolmogorov flow exhibits inverse energy cascade,  contrary to the forward energy transfer observed in three-dimensional hydrodynamic turbulence. 
 
 In the next section we will describe results from direct numerical simulation and compare them with the results of the low-dimensional model.
 
 \section{Comparison with Direct Numerical Simulations}\label{sec:level5}
 
 In this section we will compare the results of direct numerical simulation (DNS) of the Kolmogrov flow with the model results. We numerically solve  Eqs.~(\ref{eq:eom1}, \ref{eq:incompres1}) using pseudospectral method in the domain $[0, 2\pi/\alpha]\times[0, 2\pi]$ with periodic boundary conditions on all sides. We  discretize the domain into $64^2$ uniform grid points. We start the simulation with initial condition,  $\{ u_1(\mathbf{k}),  u_1(\mathbf{p}), u_1(\mathbf{q}), u_1(\mathbf{s}) \} = \{-0.01 ,0.01, -0.01(\sqrt{1+\alpha^{2}}/\alpha), 0.01(\sqrt{1+\alpha^{2}}/\alpha) \}$, with negative wavenumbers modes given by the corresponding complex conjugates.   The rest of the modes are zeros.   We employ RK2 (second order Runge-Kutta) scheme  for time advancement with fixed time step $ dt = 0.01 $.  We employ $2/3$  rule for dealiasing.
     
      \begin{table}
         \caption{ For $\alpha =0.7$ and $R=4.61$, the relative  amplitudes of the modes of low-dimensional model (LDM), as well as the relative amplitudes of the dominant modes of DNS.  The total energy of the DNS is 0.188, and that for LDM is 0.18.   The amplitudes are for the steady state at $ t=440 $.  The table does not include the $ -\textbf{k} $ modes that contain the remaining 50\% of the total energy. }
         \vspace{10pt}
         \begingroup
         \setlength{\tabcolsep}{7pt} 
         \renewcommand{\arraystretch}{1.5} 
         
         \begin{tabular}{c c c}
             \hline
             \hline
             $\textbf{k} = (k_x, k_y)$ & $E({\textbf{k}})/E(\%)$ & $E({\textbf{k}})/E(\%)$\\
             {} & (DNS) & (LDM) \\
             \hline
             ($\alpha , 0$) & $16.035$ & $15.975$ \\
             ($0,1$) & $29.459$ & $28.563$ \\
             ($\alpha,1$) & $2.197$ & $2.731$ \\
             ($\alpha,-1$) & $2.197$ & $2.731$ \\
             ($2\alpha,1$) & $0.053$ & -\\
             ($-2\alpha,1$) & $0.053$ & -\\
             ($2\alpha,2$) & $0.002$ & -\\
             ($3\alpha,1$) & $0.001$ & -\\
             ($\alpha,2$) & $0.001$ & -\\ 
             \hline
         \end{tabular}
         \endgroup
         \label{tab:amp_DNS}
     \end{table} 
  
 We perform the $30$ runs for different values of $\{\alpha, R\}$, which are displayed in Fig.~\ref{fig:neutral_curve} as green circles and blue triangles.  We observe that all the simulations reach steady solutions, which are either laminar solution ($ S_0 $, green dots) or vortex solution ($ S_1 $ or $ S_2 $, blue triangles).     Note that the two sets of simulations are nearly separated by the $ R=R_c $ curve, which is the red curve in Fig.~\ref{fig:neutral_curve}.  These observations indicate that our low-dimensional model captures the DNS results very well for the parameters of Fig.~\ref{fig:neutral_curve}.
 
 The dominant Fourier modes of the DNS  are the same as those of the low-dimensional model.  The other modes have much small magnitudes.   The flow profiles of the DNS and the model are very similar, consistent with the above observations.    For example,  for the parameter values, $ \alpha =0.7 $ and $ R=4.61 $, the steady-state flow profiles of the low-dimensional model and DNS exhibited in Figure~\ref{fig:flow_profile} are very similar.  For the same parameter values, the steady-state values of the dominant Fourier modes for the  DNS and the low-dimensional model are quite close to each other (see Table~\ref{tab:amp_DNS}).  For the DNS,  the nine modes (along with their complex conjugates) listed in Table~\ref{tab:amp_DNS} contain nearly all of the total energy of the system. 
 
 \begin{figure}[htbp]
     \centering
     \includegraphics[width=0.9 \linewidth]{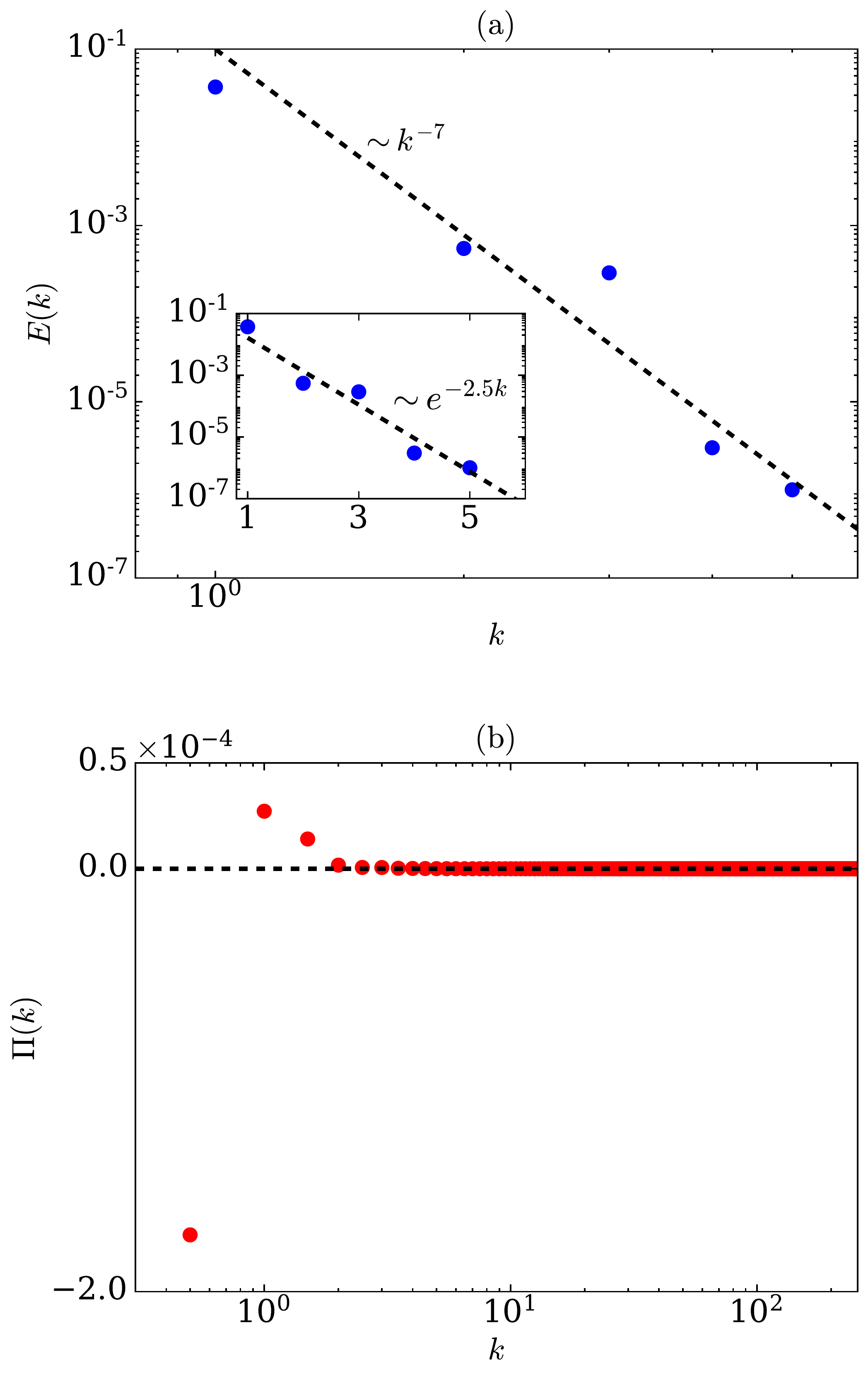}
     \caption{(color online) For DNS with $\alpha=0.7$ and $R=200$: (a) the energy spectrum $ E(k)$  and (b) energy flux $ \Pi(k) $.   Both power law ($ k^{-7} $) and exponenital function ($ \exp(-2.5k) $) [see inset of (a)] fit with the numerical data reasonably well.    $ \Pi(k) $ is negative for the lowest wavenumber, indicating inverse cascade.  Also, $ \Pi(k) $ is negligible for $ k>3 $ due to the dominance of small wavenumber modes. }
     \label{fig:Ek_Pik}
 \end{figure}

In our DNS, we do not observe  solutions other than $ S_0 $, $ S_1 $, and $ S_2 $.  That is, we do not observe any secondary bifurcation in our simulations.  The  DNS  for $ \alpha=0.7 $ and $ R=200 $ too exhibits a steady vortical flow structure as in Fig.~\ref{fig:flow_profile}, thus indicating absence of a secondary bifurcation.   Note, however, that \citet{Okamoto:JJIAM1993} had predicted secondary bifurcations for $ \alpha= 0.98$, as well as on the unstable branch for  $ \alpha=0.35 $.  These are specialized cases that require special initial conditions and careful time-advancing of the DNS; hence, this investigation is deferred for future. 

For the computation of the energy spectrum and flux, we performed a DNS for $\alpha=0.7$ and $R=200$ on a relatively higher resolution of $512^2$.   We obtain a steady flow at $t=1300$; at this time, the energy spectrum $ E(k) $ is very steep.  Steep power law of $ k^{-7} $ provides a reasonable fit to the energy spectrum, which is consistent with the predictions of \citet{Okamoto:JDDE1996}.  We also remark that the exponential function $ \exp(-2.5k) $ too provides a reasonable fit to the spectrum; this result is consistent with the arguments that the low-dimensional systems and the dissipation-range of turbulent flows exhibit exponential spectrum\cite{Paul:arxiv2009,Bershadskii:PF2008}.    Note that \citet{Zhang:POF2019} obtained similar scaling in their simulation of the Kolmogorov flow.  See Fig.~\ref{fig:Ek_Pik}(a) for an illustration.

We also compute the energy flux $ \Pi(k) $ for the same run.  The energy flux is negative for the smallest wavenumber sphere of radius 0.5, indicating an inverse cascade of energy (see Fig.~\ref{fig:Ek_Pik}(b)).  The simulation result is close to the model result, that is, the energy transfer from $ u_1(-\alpha, -1) $ to $ u_1(\alpha, 0) $ shown in Fig.~\ref{fig:triad_diagram} and discussed in Sec.~\ref{sec:level4}.  In addition, $ \Pi(k) $ falls sharply.  Thus, both the energy spectrum and flux support earlier observations that only small wavenumber modes are active in the Kolmogorov flow.  For example, see Table~\ref{tab:amp_DNS}.

We conclude in the next section. 

\vspace{1cm}
\section{Discussions and Conclusions}\label{sec:level6}
In this paper, we present a low-dimensional model that captures the essential features of the Kolmogorov flow.  The Fourier components are in the Craya-Herring basis.   We identify the fixed points of the system, and show that the system bifurcates from the laminar solution to a new solution with vortex structure.  These solutions are consistent with earlier works based on analytical, numerical, and experimental tools. In addition, we perform direct numerical simulation (DNS)  of the Kolmogorov flow that exhibits similar results as the low-dimensional  model. 

Our low-dimensional model captures the critical Reynolds number of the Kolmogorov flow. The model predicts that the new vortex solution remains stable beyond $R > R_c$.  The critical Reynolds number $R_c$ increases monotonically with  $\alpha$, with   $R_c \rightarrow \sqrt{2}$ as $\alpha\rightarrow0$, and $R_{c}\rightarrow \infty$ as $\alpha\rightarrow1$.  But between these two limits, the model prediction of  $R_c$  is marginally lower than those computed using models containing a larger number of Fourier modes~\cite{Okamoto:JJIAM1993, Nagatou:JCAM2004}.  Using energy transfers, we show that in the Kolmogorov flow, the energy flows from intermediate scales to large scales; this is contrary to the forward energy transfers in Kolmogorov's theory of turbulence. Thus, our model captures essential aspects of the primary bifurcation of the Kolmogorov flow, and its results are consistent with earlier models.    

Our DNS results are very similar to those of the low-dimensional model.  For example, the flow patterns and the dominants modes of DNS are close to those of the low-dimensional model.  Both  DNS and the model do not exhibit any secondary bifurcation, indicating the robustness of the low-dimensional model.  It is interesting to note that the six-mode dynamo model of \citet{Verma:PRE2008} showed very similar bifurcation, as described in this paper.   It is possible that the  Kolmogorov flow with forcing at larger wavenumbers ($ k_f > 1 $) may exhibit secondary bifurcation.

There are certain discrepancies between the predictions of our model and those of earlier models.  As shown by \citet{Okamoto:JJIAM1993}, we expect secondary bifurcations for $ \alpha$ very close to unity, as well as on the unstable branch for other  $ \alpha$'s.   A verification of  \citet{Okamoto:JJIAM1993}'s predictions on secondary bifurcations using DNS requires major fine-tuning of the initial conditions and the DNS, and it is planned for the future.  Also, our model does not capture several oscillatory solutions predicted by \citet{Sivashinsky:PD1985}.  These issues need to be explored in the future.

  In summary, our four-model model  captures many of its interesting features of the Kolmogorov flow.  It also opens avenues for further explorations of the Kolmogorov flow with $ k_f >1  $ and large Reynolds numbers.   



\section{Acknowledgements}
We thank Roshan Samuel, Shashwat Bhattacharya, Mohammad Anas, Narendra Pratap, Akanksha Gupta, Shadab Alam, and Manohar Sharma for useful discussions. This work was supported by the research grant 6104-1 from Indo-French Centre for the Promotion of Advanced Research (IFCPAR/CEFIPRA). Soumyadeep Chatterjee is supported by INSPIRE fellowship (IF180094) of Department of Science \& Technology, India.  


\section{Data Availability}
The data that support the findings of this study are available from the corresponding author upon reasonable request.


\section{References}


\providecommand{\noopsort}[1]{}\providecommand{\singleletter}[1]{#1}

\end{document}